# Towards radiation pressure acceleration of protons using linearly polarized ultrashort petawatt laser pulses

I Jong Kim[1,3], Ki Hong Pae[1], Chul Min Kim[1,3], Hyung Taek Kim[1,3], Jae Hee Sung[1,3], Seong Ku Lee[1,3],

Tae Jun Yu[1,3], Il Woo Choi[1,3], Chang-Lyoul Lee[1], Kee Hwan Nam[1], Peter V. Nickles[2], Tae Moon Jeong[1,3]*,

and Jongmin Lee[1]**

[1]Advanced Photonics Research Institute, GIST, Gwangju 500-712, Korea

[2]WCU, Department of Nanobio Materials and Electronics, GIST, Gwangju 500-712, Korea

[3]Center for Relativistic Laser Science, Institute for Basic Science GIST-campus, Gwangju 500-712, Korea

jeongtm@gist.ac.kr* and leejm@gist.ac.kr**

Particle acceleration using ultraintense, ultrashort laser pulses is one of the most attractive topics in relativistic laser-plasma research. We report proton/ion acceleration in the intensity range of $5\times10^{19}$ W/cm$^2$ to $3.3\times10^{20}$ W/cm$^2$ by irradiating linearly polarized, 30-fs, 1-PW laser pulses on 10- to 100-nm-thick polymer targets. The proton energy scaling with respect to the intensity and target thickness was examined. The experiments demonstrated, for the first time with linearly polarized light, a transition from the target normal sheath acceleration to radiation pressure acceleration and showed a maximum proton energy of 45 MeV when a 10-nm-thick target was irradiated by a laser intensity of $3.3\times10^{20}$ W/cm$^2$. The experimental results were further supported by two- and three-dimensional particle-in-cell simulations. Based on the deduced proton energy scaling, proton beams having an energy of ~ 200 MeV should be feasible at a laser intensity of $1.5\times10^{21}$ W/cm$^2$.

Charged particle acceleration using ultraintense, ultrashort laser pulses is one of the most attractive topics not only in high field science [1-5] but also for many applications. One of the most challenging applications driving recent activities is the laser-based proton/ion acceleration for medical application in cancer therapy [6]. At currently available laser intensities, only indirect proton/ion acceleration is possible and it relies on the spatial and temporal dynamics of electrons that are directly accelerated by the laser pulse. The electric field induced by the charge separation between the remaining heavier ions in the target and the fast moving electrons from the target forms the driving force for accelerating protons. In general, the electron properties depend on the laser and target parameters (e.g., peak intensity, contrast, polarization, spatial and temporal shape, target density, thickness, geometry and atomic numbers of the constituents) [7,8] and the electron energy distribution is defined by a high-temperature, broad thermal distribution and a low-temperature, narrow collective or coherent distribution. Accordingly, the proton/ion acceleration mechanism and their maximum energy depend on the properties of the electrons that contribute to the proton/ion acceleration.

Protons/ions acceleration by thermal electrons is known as target normal sheath acceleration (TNSA) [9-11] and is valid for solid targets with a wide thickness range, hundreds of nanometers to a few micrometers. In the TNSA scheme, the maximum proton energy ($E_{p,max}$) is determined by the thermal electron temperature ($T_e = m_e c^2 [(1+a_0^2)^{1/2}-1]$) [12], where $m_e$ is the electron mass, $c$ is the speed of light, and $a_0$ is the normalized vector potential. The proton energy should thus scale with the peak intensity in the form of $I^{1/2}$ [7]. Particle-in-cell (PIC) simulations have suggested that a maximum proton energy of > 300 MeV should be possible at a peak intensity of $1\times10^{22}$ W/cm$^2$, which is the highest laser intensity demonstrated [13]. Recently, a maximum proton energy of 40 MeV has been reported using μm-thick metal foils irradiated by 40-fs, 7.5-J laser pulses at $1\times10^{21}$ W/cm$^2$ [14], but no energy scaling result was shown in that report.

Protons can be accelerated directly by a laser pulse at intensities of over $10^{24}$ W/cm$^2$. However, with a circularly polarized (CP) laser, the required intensity for protons and ions acceleration can be greatly reduced by the formation of an electrostatic field due to space charge separation between ions and compressed electrons [15]. The use of a nanometer-scale ultrathin target with a CP laser pulse reduced the required intensity further to the level of $10^{21}$ W/cm$^2$ [16]. This acceleration process is called light-sail radiation pressure acceleration (RPA) mechanism. According to the RPA model, most electrons in an ultrathin target irradiated by a CP laser pulse are collectively accelerated by the radiation pressure and the maximum proton energy linearly scales with the laser intensity of *I*. The acceleration characteristics can be interpreted in a general way using the normalized areal density $\sigma_{opt} = (n_e/n_c)(d/\lambda_L)$, where $n_e$ is the electron number density, $n_c$ is the critical density of the plasma, $d$ is the target thickness, and $\lambda_L$ is the laser wavelength. Optimal conditions are obtained when the areal density is approximately equal to $a_0$. An RPA-like scheme has been demonstrated using a $5 \times 10^{19}$ W/cm$^2$ CP laser pulse [17] in which a quasi-monoenergetic feature in the C$^{6+}$ ion spectrum and a significant reduction in thermal electron component were observed. However, due to the limited laser intensity, it was not possible to confirm the energy scaling and monoenergetic feature in the proton spectrum. But, recent observation for a narrow-band feature in the proton/ion energy spectrum was reported from an experiment using sub-picosecond (700fs), CP and linearly polarized (LP), laser pulses with a maximum intensity of $3\times10^{20}$ W/cm$^2$ [18]. Alternative RPA schemes such as the hybrid RPA-TNSA [19, 20] and leaky light-sail RPA [21] have been proposed for stable generation of high energy mono-energetic proton/ion beams even at lower intensities ($10^{20} - 10^{21}$ W/cm$^2$) and/or with linear polarization.

In this article, we report on the experimental and simulation results for proton/ion acceleration from ultrathin polymer target using 30 fs, LP, petawatt (PW) laser pulses. The aim of the experiments was to study, by measuring the proton energy scaling with respect to the laser intensity for different target thicknesses, the proton acceleration mechanism in the intensity range of $5 \times 10^{19}$ W/cm$^2$ to $3.3 \times 10^{20}$ W/cm$^2$ and to find out the conditions for increasing the maximum proton energy with LP laser pulses. As a result, we experimentally showed, for the first time, the transition of acceleration mechanism from the TNSA to RPA regime when a target with a thickness below 30 nm was irradiated by an ultraintense laser pulse. Two-dimensional (2D) and three-dimensional (3D) particle-in-cell (PIC) simulations were carried out under the same physical conditions to interpret the experimental results. Simulation results not only agreed well with the energy scaling regarding the laser intensity and target thickness, but also reproduced the structure and temporal evolution of proton spectra. Interesting features indicating the action of RPA such as linear energy scaling, quasi-monoenergeticity, low electron temperature, and strong longitudinal electrostatic field by an electron-proton double layer are presented below, along with detailed numerical results and the interpretation. Finally, the importance of this article and the perspective of proton energy enhancement are described in conclusion.

Linearly-polarized, 30-fs, 27-J laser pulses from the 1-PW, Ti:sapphire laser system (PULSER beamline I) at the Advanced Photonics Research Institute (APRI) were delivered to a double plasma mirror (DPM) system to enhance the temporal contrast ratios (Methods). The overall output performances of the PW laser system are described elsewhere [22]. The DPM system had the same geometry as that used for the 100 TW laser system except for the size [23]. After the DPM system, 30-fs, 8.3-J s-polarized laser pulses were focused onto ultrathin polymer targets (F8BT; Methods) with thicknesses of 10, 20, 30, 50, 70, and 100 nm. The typical spot size in the focal plane was 5.8 μm at FWHM, resulting in the maximum intensity of $3.3 \times 10^{20}$ W/cm$^2$ ($a_0 = 12.4$). The laser pulses were incident at an angle of 7° with respect to the target normal to avoid retro-reflection from the plasma on the target. During the experiment, the laser intensity was varied from $5 \times 10^{19}$ W/cm$^2$ ($a_0 = 4.8$) to $3.3 \times 10^{20}$ W/cm$^2$ ($a_0 = 12.4$) by turning on pump-laser-beam lines sequentially without degrading the quality of focal spot. The proton and ion energy spectra were measured by a Thomson parabola equipped with a micro-channel plate (MCP) and an intensified CCD (ICCD, Princeton Instruments, PI-MAX), and the absolute number calibration for the proton energy spectra was done by installing striped CR-39 track detectors in front of the MCP device [24]. The detailed information on the laser system, target, electron and proton diagnostics is given in the Methods and the Supplementary Information.

The proton and carbon ion energy spectra obtained from a 10-nm-thick target irradiated by a laser pulse with an intensity of $3.3 \times 10^{20}$ W/cm$^2$ is shown in Fig. 1a. The maximum proton and C$^{6+}$ ion energies were 45 MeV and 164 MeV, respectively, which are the highest values ever reported from acceleration experiments using tens-of-fs high-power laser pulses. For the 100-nm-thick target, maximum energies of 18 MeV and 60 MeV were observed for proton and C$^{6+}$ ion, respectively. Figure 1b shows the maximum proton energies obtained at different target thicknesses and intensities. It is clearly visible that, at a given intensity, the maximum proton/ion energy increases as the target thickness decreases. Thus, for the comparison with CP laser cases, it might be necessary to find out the optimum thickness for LP laser pulses by using targets thinner than 10 nm, which were not available.

Figures 2a and b show the proton and C$^{6+}$ ion energy spectra, respectively, obtained from the 10-nm- and 100-nm-thick targets irradiated at the highest intensity of $3.3 \times 10^{20}$ W/cm$^2$. The proton spectrum for the 10-nm target showed a broad and modulated profile. PIC simulations have shown that a broad spectral profile results from the temporal evolution of electrons and protons during the acceleration process and that the modulation is related to the Rayleigh–Taylor (RT) instability (See the details in the simulation part). Interestingly, for the 10-nm target, the quausi-monoenergetic peaks (at 65, 91, and 120 MeV) in the C$^{6+}$ ion spectrum are clearly visible in Fig. 2b and are similar to those shown in ref. 17. This feature was observed in the successive measurements and may indicate the evidence of RPA mechanism in the acceleration process under these conditions (a 10- to 30-nm-thick target irradiated by a $3.3 \times 10^{20}$ W/cm$^2$ LP laser pulse). For the 100-nm-thick target, however, the C$^{6+}$ ion spectrum showed exponential decay, which is a typical feature of the TNSA mechanism. Despite the quasi-monoenergetic feature in the ion spectrum, no obvious monoenergetic structure was observed in the proton energy spectrum even at the highest intensity ($3.3 \times 10^{20}$ W/cm$^2$). This phenomenon was also reproduced by 2D and 3D PIC simulations.

The dependence of the maximum proton energy on the laser intensity for the 10-, 20-, and 30-nm targets shown in Fig. 3a reveals the RPA features in the proton acceleration (See also Supplementary Fig. S6 for the comparison). The maximum proton energy at each intensity is the average of three laser shots, and the error bars in the figure denote the standard deviations of the maximum proton energy and laser peak intensity. For these targets, the proton energy scaling shows a very important feature: at lower intensities, the maximum proton energy increased with the intensity in a form of $I^{1/2}$, and then linearly increased with the intensity after a certain thickness-dependent-transition intensity. This behavior is explained by the transition from TNSA to RPA regimes. The transition occurred at an intensity of approximately $1.75 \times 10^{20}$ W/cm$^2$ for the 10-nm target, and the transition intensity increased by about $1.5 \times 10^{19}$ W/cm$^2$ per 10 nm; a thicker target showed a higher transition intensity due to its higher areal density. For 20-nm and 30-nm targets, the energy scaling varies abruptly from $I^{1/2}$ to $I^1$, which implies that a small difference in the laser intensity may cause the change in the dominant acceleration mechanism. As shown in Fig. 3b, the PIC simulations are in good agreement with the experimental results. A similar trend was seen for the energy scaling of the C$^{6+}$ ions (Fig. S5 in the Supplementary Information).

The TNSA to RPA transition is also supported by the electron energy spectra (measured simultaneously), as shown in Fig. 4. For the 100-nm-thick target, the electron spectrum is broad and dominated by thermal electrons with an electron temperature of $T_e = 1.6$ MeV. Thermal electrons are responsible for proton acceleration from targets thicker than 50 nm via the TNSA mechanism. For the RPA with thicker targets, a higher laser intensity is required to expel most of the electrons in a collective way from a target. In contrast, for the 10-nm-thick target, two electron components were clearly visible in the spectrum: collective electrons with $T_e = 0.4$ MeV and thermal electrons with $T_e = 3.6$ MeV. Thus, it can be assumed that both collective and thermal electrons are responsible for the proton/ion acceleration, resulting in the coexistence of TNSA and RPA mechanisms. In such a condition, the collective electron bunch produces the high-energy protons and quasi-monoenergetic $C^{6+}$ ions, while the thermal electrons accelerate relatively low energy protons with a broad spectral profile. The changes in proton energy scaling and electron spectrum with respect to the intensity are experimental evidences for a transition from the TNSA to RPA regime.

To visualize the temporal evolution of electrons, protons, and ions during the acceleration process and to explain the experimental results, we carried out 2D and 3D PIC simulations showing the dynamics of the electrons, protons, and ions for both 10-nm- and 30-nm targets irradiated by an ultraintense, ultrashort laser pulse. The peak intensity was varied from $5 \times 10^{19}$ W/cm$^2$ to $3.5 \times 10^{20}$ W/cm$^2$; additional details are given in the Supplementary Information. The APRI laser plasma simulator (ALPS) code [25 - 27] was used.

Figure 5 shows the 3D PIC simulation results for the 10-nm thin target irradiated at $3 \times 10^{20}$ W/cm$^2$. The temporal evolutions of the electron and proton number densities for the 10-nm target are shown in Figs. 5a and b. The acceleration dynamics can be described as follows: at 24 fs after the arrival of a laser pulse, a portion of electrons, heated by the oscillating component in the ponderomotive force, leave the target with a broad energy distribution (thermal electrons) in both forward and backward directions. At 36 fs, two different proton beam components appear in the interaction region: one due to the thermal electrons and the other due to the collective electrons. The thermal electrons drag protons in the forward and backward directions, and these protons obtain a broad momentum distribution (TNSA). The collective electrons pushed by radiation pressure contribute to form an electron-proton double layer [16, 20] and drag protons to have a narrow momentum distribution in the direction of the laser beam propagation. At the same time, a return current is formed, and the Rayleigh-Taylor-like instability (RTI) [28] begins to grow, resulting in the spectral modulation of the proton energy spectrum. The proton acceleration continues as long as the laser pulse intensity is high enough to maintain the double layer structure. At 60 fs, the collective electron layer smears out, and the proton acceleration is weakened due to the spatial broadening of the electrons. According to the previous research [20], with a spatially trapezoidal laser pulse at a higher intensity of $10^{21}$ W/cm$^2$, the structure can be maintained for a longer time to produce GeV C$^{6+}$ ions with quasi-monoenergetic spectrum. The protons accelerated by the collective electrons is faster than those accelerated by thermal electrons, and eventually, with an s-polarized laser field, the RTI results in the fine structure in the electron and proton density maps. In contrast, due to the high mass of carbon ions, the C$^{6+}$ ions are not heavily influenced by thermal electrons, resulting in a quasi-monoenergetic spectrum. For a 30-nm target irradiated by a $1 \times 10^{20}$ W/cm$^2$ laser pulse, the collective electrons produced by the high-intensity part of the laser pulse are not sufficient, and proton acceleration is dominated by the TNSA mechanism, as depicted in Supplementary Fig. S7.

Figure 5c shows a temporal evolution of the protons in momentum space in which the protons accelerated by collective electrons are faster than those accelerated by thermal electrons, leading to the linear energy scaling of the maximum proton energy. The temporal evolutions of the electric fields at two different conditions (30-nm target at $1 \times 10^{20}$ W/cm$^2$ and 10-nm target at $3 \times 10^{20}$ W/cm$^2$) are shown in Fig. 5d. A strong longitudinal electrostatic field due to the electron-proton double layer is formed at an intensity of $3 \times 10^{20}$ W/cm$^2$ and maintains until the collective electrons broaden. As shown in Figs. 5c and d, thermal electrons appear in the early stage of interaction and contribute to the proton and ion acceleration during the whole process, while collective electrons appear in the high-intensity part and contribute to the proton and ion acceleration only in a specific time period. Consequently, we can conclude that the collective electrons are responsible for the RPA features such as the linear scaling of maximum proton energy and quasi-monenergetic structure in the C$^{6+}$ ion spectrum.

**Conclusion**

The experiments performed with LP 30-fs, 1-PW laser pulses have produced the maximum proton and $C^{6+}$ ion energies of 45 MeV and 164 MeV, respectively, at a laser intensity of $3.3 \times 10^{20}$ W/cm$^2$. The change in the energy scaling indicated a transition from TNSA to RPA with targets of which thickness is below 30 nm. RPA features of the accelerated protons and ions were also supported by the electron spectrum measured simultaneously. The 2D- and 3D-PIC simulations gave a detailed understanding of the acceleration dynamics, and produced the spectra and the energy scaling that agreed well with the experimental results.

Assuming the validity of the measured RPA scaling for the 10 nm polymer target, a proton energy of 190 MeV and a $C^{6+}$ ion energy of 732 MeV (See Figure 6) should be possible at an intensity of $1.5 \times 10^{21}$ W/cm$^2$, which is expected to be available soon [29]. Such a high value of particle energy will make a breakthrough for the cancer treatment using laser-accelerated protons and ions, and thus further experimental studies on the RPA scaling at higher laser intensity are of crucial importance. Experiments at intensities beyond $1 \times 10^{21}$ W/cm$^2$ will bring more insight in the dynamics and validity of an optimal RPA regime characterized by a linear intensity scaling of proton or ion energy.

**Methods**

**Laser system.** The experiments were performed at the PW Ultrashort Laser Source for Extreme science Research I (PULSER I) of the APRI, GIST. The PULSER I is a CPA Ti:sapphire laser delivering after the compressor up to 1 PW in a 30 J, 30 fs pulse at 800 nm wavelength. This laser pulse with a 200-mm beam diameter was propagated to a double plasma mirror (DPM) system to ensure a high pulse contrast. (see Supplementary Figure S1). In the DPM system, the laser pulse is focused with an F/10 off-axis parabolic (OAP) mirror, and its focal plane is located at the midpoint between antireflection-coated two plasma mirrors. Half-wave plate is installed to reduce the resonance absorption [30] of laser pulse on DPM by switching the laser polarization from p-polarization to s-polarization. As shown in Supplementary Figure S2, the reflectivity of DPM for s-polarized laser is

~39 % whereas that for p-polarized laser is ~16 % at the energy fluence of 127 J/cm$^2$. The total reflectivity of PW DPM system becomes 32 %, mainly due to the transmission of half-wave plate and the reflectivity of DPM. After reflection from a second plasma mirror, the laser beam is collimated by another OAP mirror. The beam size is then increased to its original diameter before the beam is directed into the target chamber.

After the DPM system, the s-polarized laser beam is focused on the target surface using an 30º OAP mirror (f/4, f = 800 mm) (see Supplementary Figure S3). The target surface is placed at an incident angle of 7° with respect to the target normal to preserve the laser front-end from amplified back reflected signal from target. Since the total reflectivity of DPM system is only 32 % with respect to the laser pulse energy, the presently available energy of laser pulse on the target is 8.3 J (0.28 PW). The measured focal spot size is a 5.4 μm and 6.1 μm FWHM spot in both the horizontal and vertical directions (see Supplementary Figure S4a), and the calculated energy concentration within the FWHM is around 50 %. Consequently, the peak intensity on the target is $3.3 \times 10^{20}$ W/cm$^2$. The temporal contrast of the laser pulse measured with a third-order cross-correlator (Amplitude Technologies, SEQUOIA) up to 500 ps before the main pulse, is shown in Supplementary Figure S4b. The measured contrast ratio with DPM system is about $3 \times 10^{-11}$ at 6 ps before the main pulse. The inset in Supplementary Figure S4b shows the temporal profiles measured with Spectral Phase Interferometry for Direct Electric-field Reconstruction (SPIDER).

**Proton/Ion diagnostics.** For the detection of the proton spectra with high energy and charge-to mass-resolution a Thomson parabola (TP) spectrometer with parallel magnetic (B = 0.4 T) and electric (E = 20 keV/cm) field was used (see Supplementary Figure S3). The parabolic ion traces are recorded using a microchannel plate (MCP) with the phosphor screen imaged to a 16-bit ICCD camera (Princeton Instruments, PI-MAX). The solid angle of the TP is $1.7 \times 10^{-8}$ steradians (sr). The calibration of the MCP-phosphor screen-ICCD system was performed by installing slotted CR-39 track detectors in front of the MCP [24]. The upper boundary for an energy error was less than 1 MeV for ~45 MeV proton, that is less than 2% and decreases with the reduction of proton energy.

**Electron Diagnostics.** For the detection of electron spectra, a magnetic electron spectrometer equipped with Fujifilm BAS-SR imaging plate was positioned behind the target at an angle of 11.5° with respect to the target normal. The magnetic field applied was B =0.52 T and the solid angle of the electron spectrometer was $1.9 \times 10^{-5}$ sr.

**Target.** The ultrathin targets were made of a conjugated polymer, poly(9,9′-dioctylfluorence-*co*-benzothiadiazole) (F8BT) [31]. The composition ratio of hydrogen to carbon was 1.2:1. The material was spin-coated onto a silicon wafer and a film thickness is in the 10–100 nm. The film was floated on water and then transferred to a special holder consisting of an array of holes to provide the freestanding foil target. The thickness and surface quality of the target were characterized using X-ray reflectivity measurements and a surface profiler respectively.

**PIC simulation.** Two- and three-dimensional particle-in-cell simulations were carried out using the fully relativistic electromagnetic code ALPS. ALPS code has already been successfully used to simulate laser driven electron [25] and proton accelerations [26] as well as to study relativistic harmonics generation via oscillatory flying mirrors [27]. The targets are modeled by cold plasmas composed of multi-species ions ($C^{6+}$ ions and protons with number density ratio of $n_C : n_H$ 1:1.5) and electrons with realistic solid density. The grid mesh size and time step were carefully chosen to resolve the electron dynamics within the relativistic collisionless skin depth ($l_s = \gamma^{1/2} c/\omega_p$) and fine structures caused by instabilities. The laser pulse is modeled using an Gaussian spatial profile (6 µm FWHM spot size) and a $\sin^2$ intensity temporal profile (30 fs FWHM duration). The laser pulse is irradiated on the target at an angle of 7° with the focal plane at the target center. In all simulations, fields at a given particle position are determined using cubic interpolation to push macro-particles and an exact charge conservation scheme [32] was used for the current density deposition.

**Acknowledgments**

This work was supported by the Ministry of Knowledge and Economy of Korea through the Ultrashort Quantum Beam Facility Program. Also this work was supported by the Research Center Program of IBS (Institute for Basic Science) in Korea and Applications of Femto-Science to Nano/bio-Technology Utilizing Ultrashort Quantum Beam Facility through a grant provided by GIST. PVN acknowledges support of the World Class University program (R31-2008-000-10026-0) grant provided by National Research Foundation (NRF) of Korea.

**Author contributions**

J. L. and T. M. J. supervised the experiment. I J. K., H. T. K. and K. H. N. carried out the experiment. J. H. S., S. K. L., and T. J. Y. developed the laser system. I. W. C and C. L. L. fabricated the target. K. H. P. and C. M. K. performed simulations and developed the theory. T. M. J., I J. K., K. H. P., C. M. K., and P. V. N. analyzed and discussed the experimental and simulation results. Finally, T. M. J., I J. K., K. H. P., C. M. K and P. V. N wrote the manuscript.

**Additional Information**

The authors declare that they have no competing financial interests. Correspondence and requests for materials should be addressed to leejm@gist.ac.kr and jeongtm@gist.ac.kr

**Figure legends**

**Figure 1. Proton and $C^{6+}$ energy measured from the Thomson parabola. a,** Energy spectra of the protons and carbon ions obtained from a 10-nm-thick target irradiated by a laser pulse with an intensity of $3.3 \times 10^{20}$ W/cm$^2$. **b,** Maximum proton energies obtained for different target thickness and intensities.

**Figure 2. Energy spectra of protons and $C^{6+}$ ions obtained from 10- and 100-nm-thick targets. a,**

Proton and **b**, $C^{6+}$ energy spectrum obtained with an intensity of $3.3 \times 10^{20}$ W/cm$^2$.

**Figure 3. Maximum proton energy as a function of the laser peak intensity for linearly polarized laser pulse. a,** Maximum proton energy for 10-, 20-, and 30-nm-thick polymer target. **b,** Comparison between the simulated and experimental results.

**Figure 4. Electron energy spectra for 10- and 100-nm targets.** For the 10 nm target, two electron components are observed: collective electrons with a temperature of $T_e = 0.4$ MeV and thermal electrons with $T_e = 3.6$ MeV. In contrast, for 100-nm target, only one electron component is observed: thermal electrons with $T_e = 1.6$ MeV.

**Figure 5. PIC simulation results.** Temporal evolutions of number density of **a,** electrons and **b,** protons from a 10-nm target irradiated by an intensity of $3.0 \times 10^{20}$ W/cm$^2$ and **c,** the corresponding proton phase space distribution. **d,** Temporal evolutions of the longitudinal electrostatic field for a 10-nm target irradiated by an intensity of $3.0 \times 10^{20}$ W/cm$^2$ and a 30-nm target irradiated by an intensity of $1.0 \times 10^{20}$ W/cm$^2$. Results are shown for t = 24, 36, 48 and 60 fs after the start of interaction.

**Figure 6. Extension of proton and $C^{6+}$ ion energy scaling. a,** 45 MeV ($H^+$) and 160 MeV ($C^{6+}$) for $3.3 \times 10^{20}$ W/cm$^2$, **b,** 128 MeV ($H^+$) and 491 MeV ($C^{6+}$) for $1.0 \times 10^{21}$ W/cm$^2$, and **c,** 190 MeV ($H^+$) and 732 MeV ($C^{6+}$) for $1.5 \times 10^{21}$ W/cm$^2$. These results were obtained under the assumption that the linear scaling of the maximum proton energy obtained from a 10-nm polymer target extends to $1.5 \times 10^{21}$ W/cm$^2$.

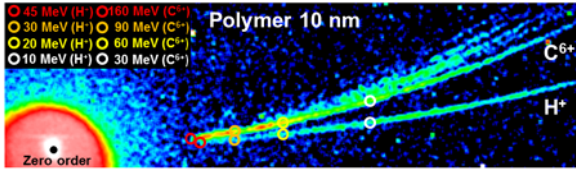 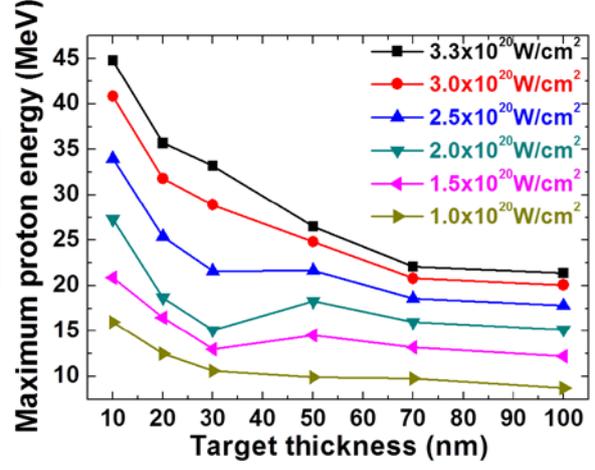

Figure 1 of 6.

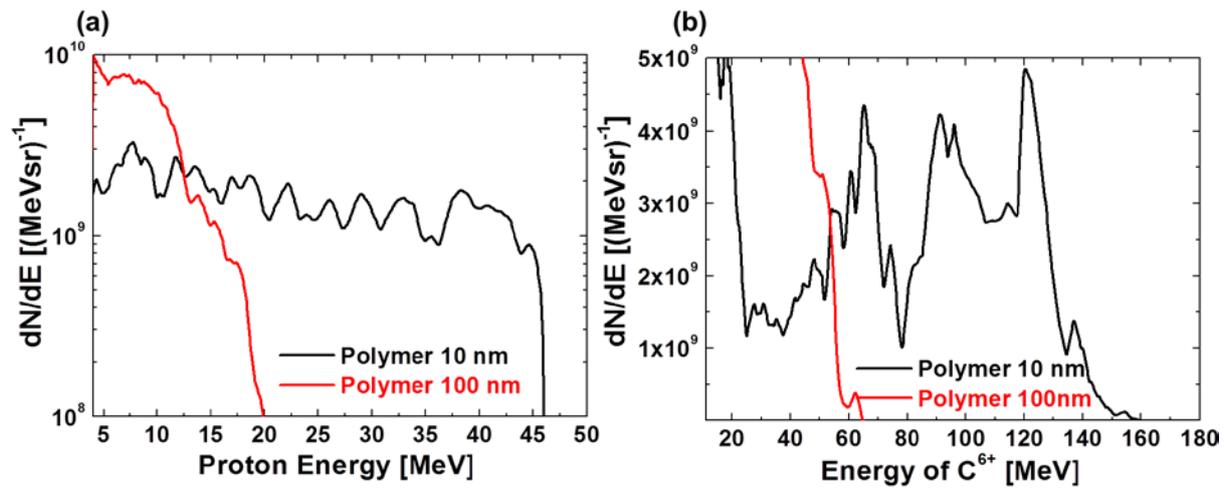

Figure 2 of 6.

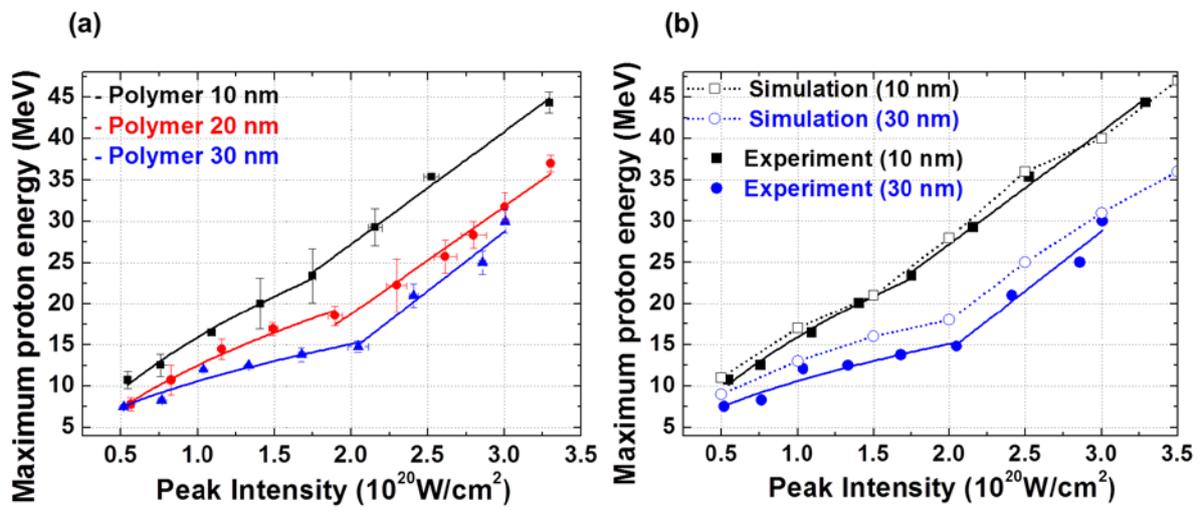

Figure 3 of 6.

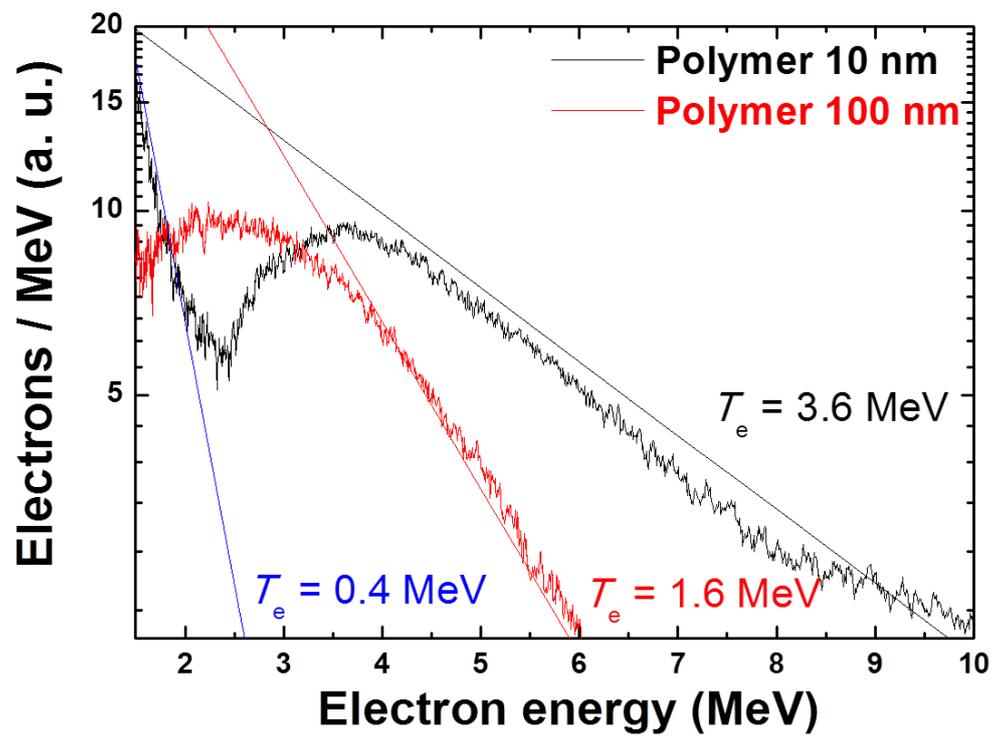

Figure 4 of 6.

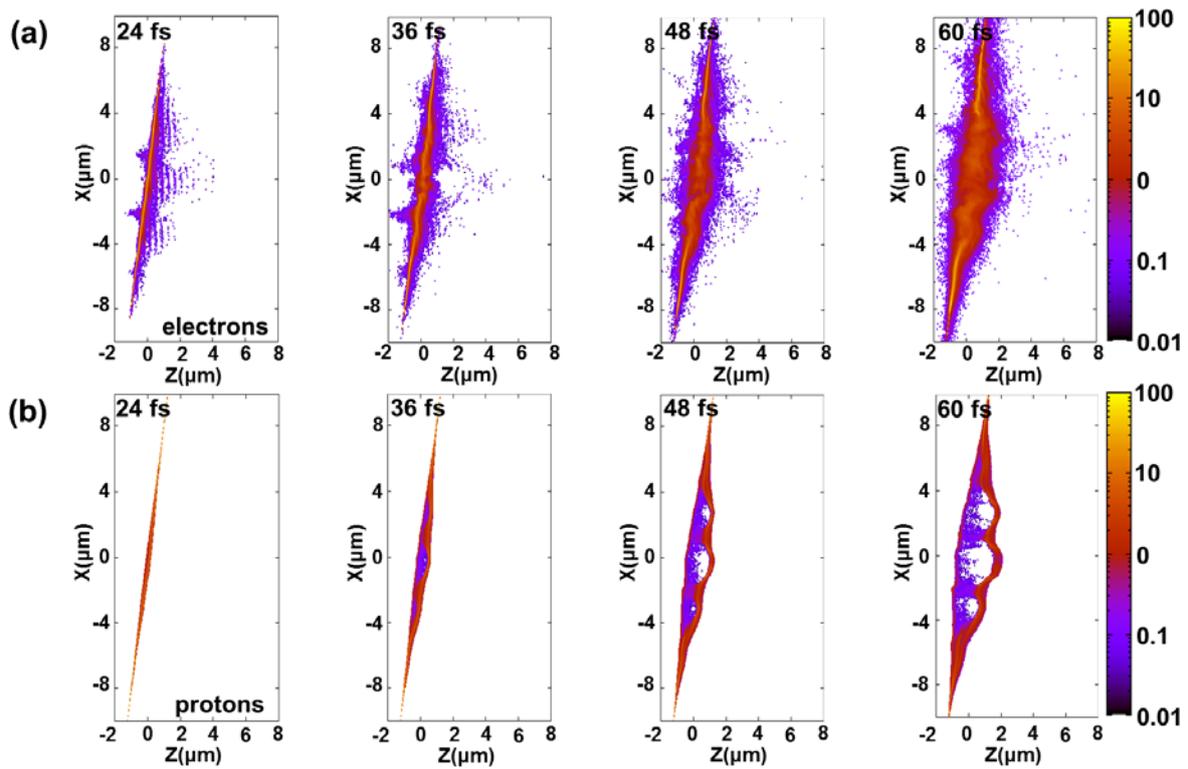

Figure 5 (a) and 5 (b) of 6.

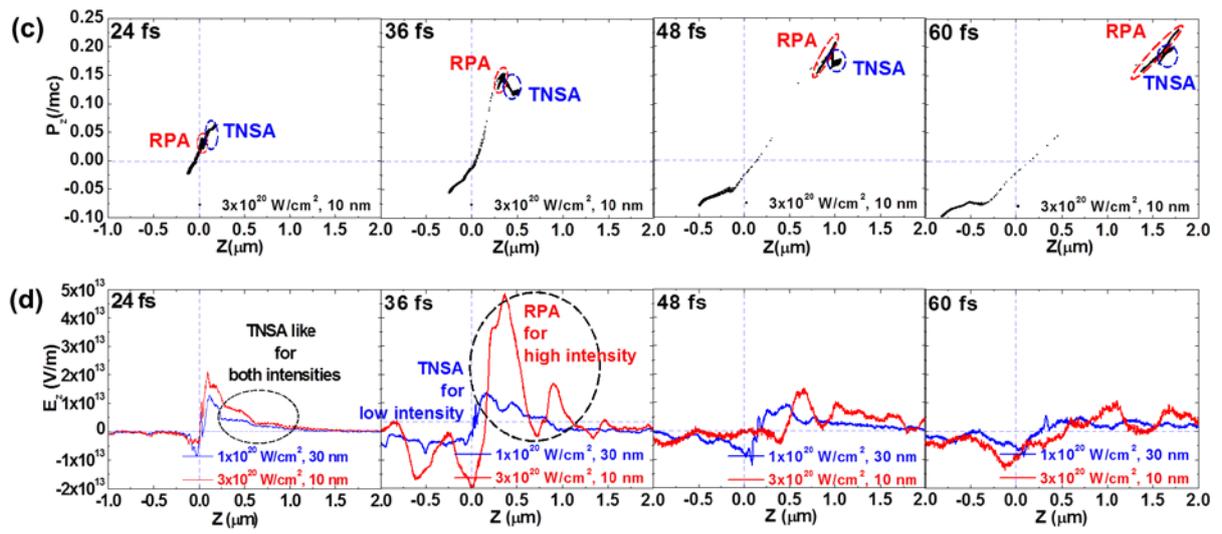

Figure 5 (c) and 5 (d) of 6.

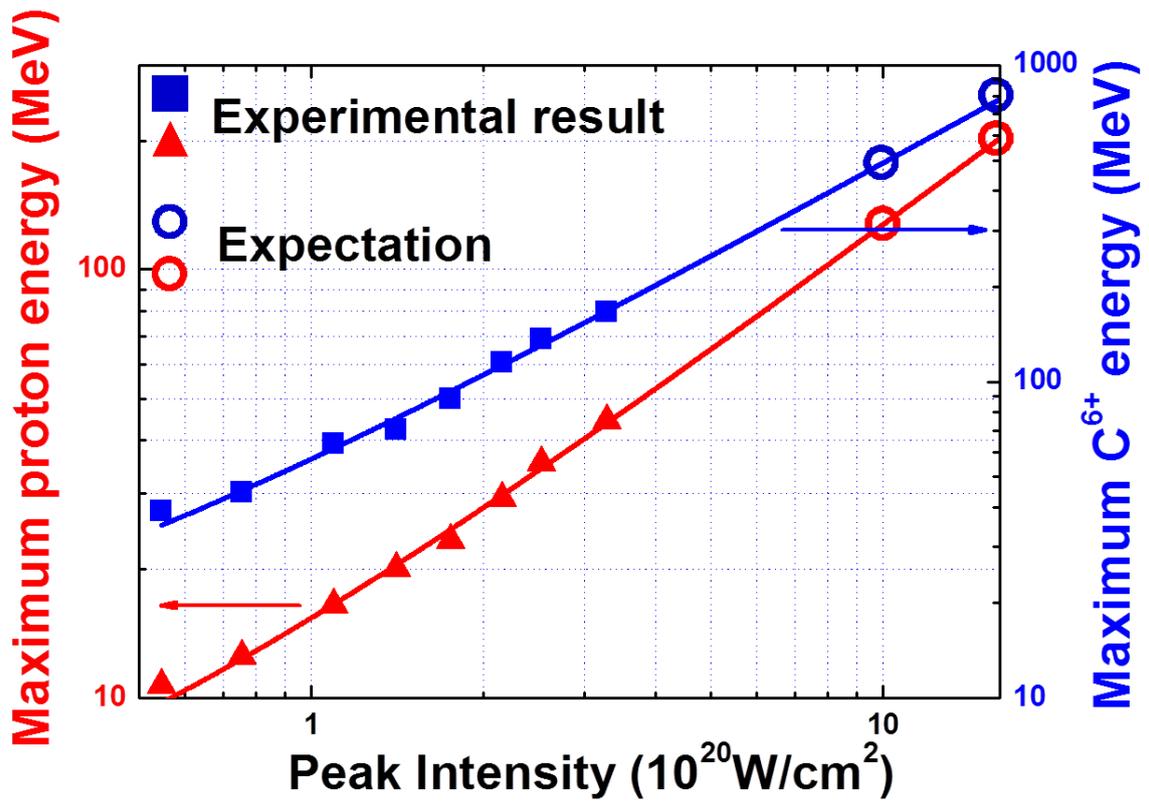

Figure 6 of 6.

# Supplementary Information

# Towards radiation pressure acceleration of protons using linearly polarized ultrashort petawatt laser pulses

I Jong Kim[1,3], Ki Hong Pae[1], Chul Min Kim[1,3], Hyung Taek Kim[1,3], Jae Hee Sung[1,3], Seong Ku Lee[1,3], Tae Jun Yu[1,3], Il Woo Choi[1,3], Chang-Lyoul Lee[1], Kee Hwan Nam[1], Peter V. Nickles[2], Tae Moon Jeong[1,3], and Jongmin Lee[1]

[1]Advanced Photonics Research Institute, GIST, Gwangju 500-712, Korea

[2]WCU, Department of Nanobio Materials and Electronics, GIST, Gwangju 500-712, Korea

[3]Center for Relativistic Laser Science, Institute for Basic Science GIST-campus, Gwangju 500-712, Korea

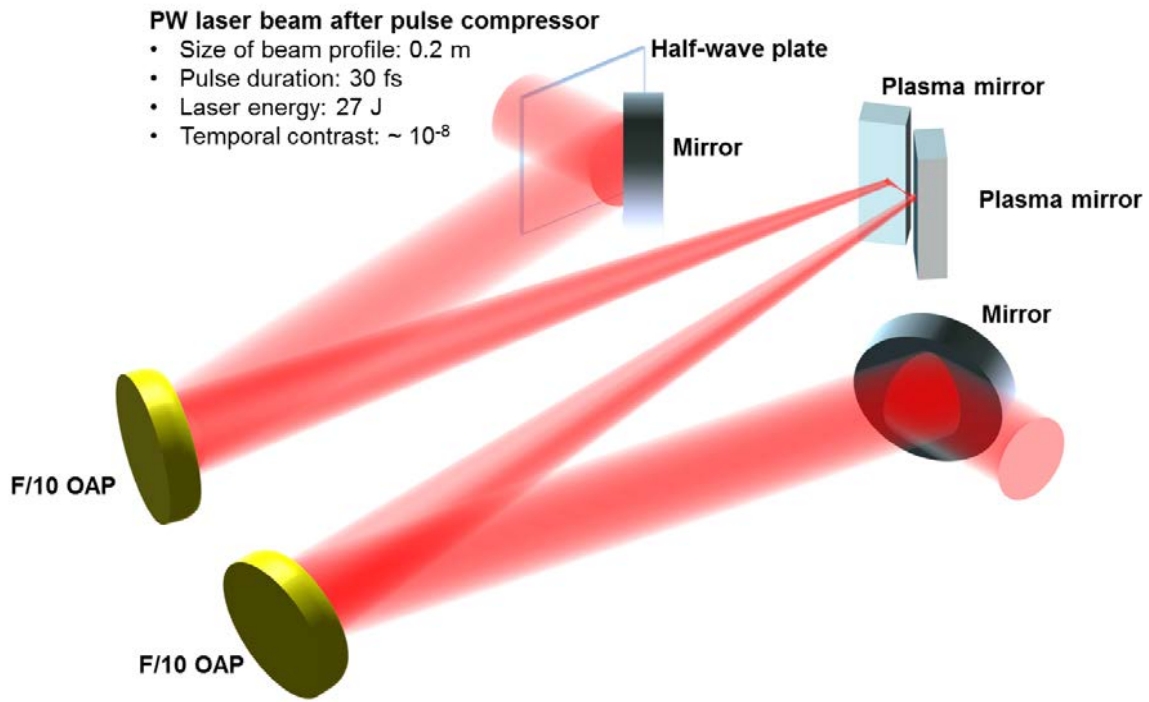

**Supplementary Figure S1: Schematic diagram of PW DPM system.** Half-wave plate is installed to reduce the resonance absorption of laser pulse on DPM by switching the laser polarization from p-polarization to s-polarization.

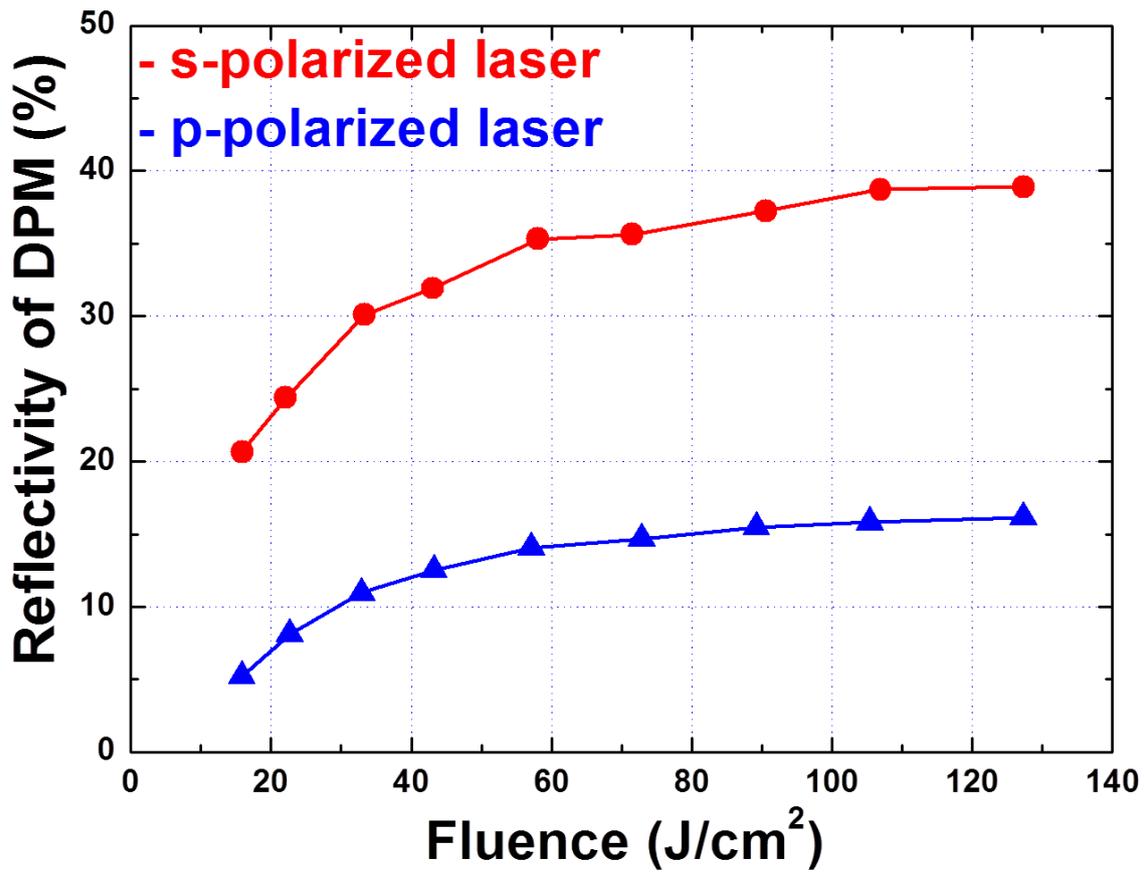

**Supplementary Figure S2: Reflectivity of DPM.** Reflectivity of DPM as a function of energy fluence incident on the first plasma mirror.

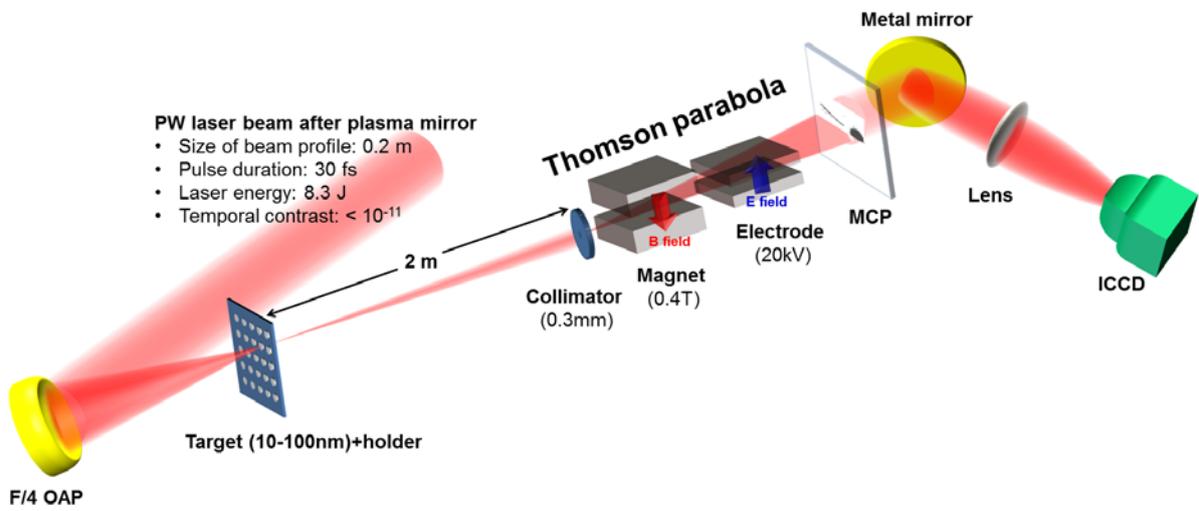

**Supplementary Figure S3: Schematic diagram of target configuration and Thomson parabola.** A short, high-intensity laser pulse is focused on an ultrathin polymer target by off-axis parabolic mirror (OAP). Ions are recorded by real-time Thomson parabola (TP) spectrometer using micro channel plate (MCP).

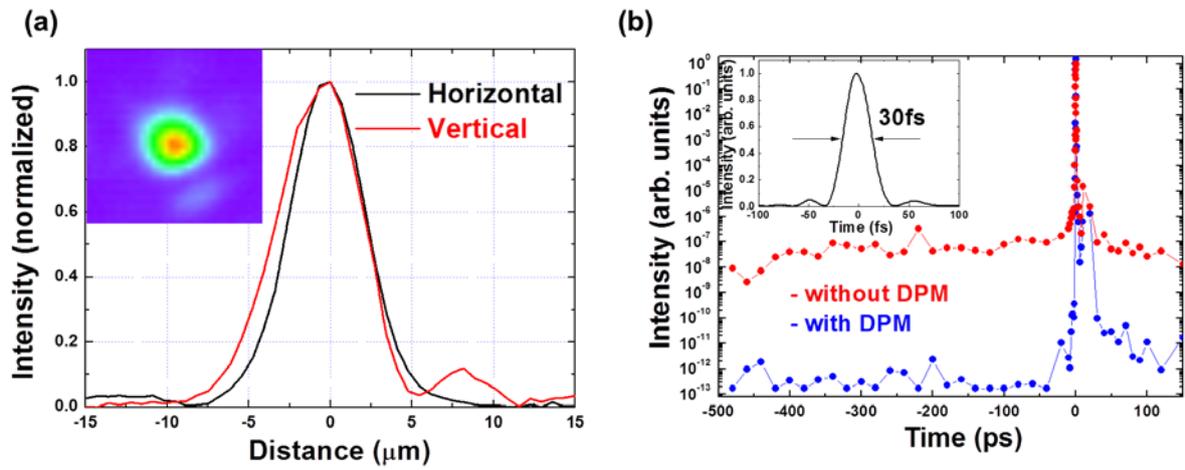

**Supplementary Figure S4: Spatio-temporal profiles of laser beam.** (a) Spatial beam profiles of a focused laser after F/4 OAP. The inset shows the focused image of laser beam. (b) Temporal profiles of the laser beam between -500 ps and +150 ps is obtained by third-order cross correlator. The inset in (b) shows the temporal profiles measured with SPIDER.

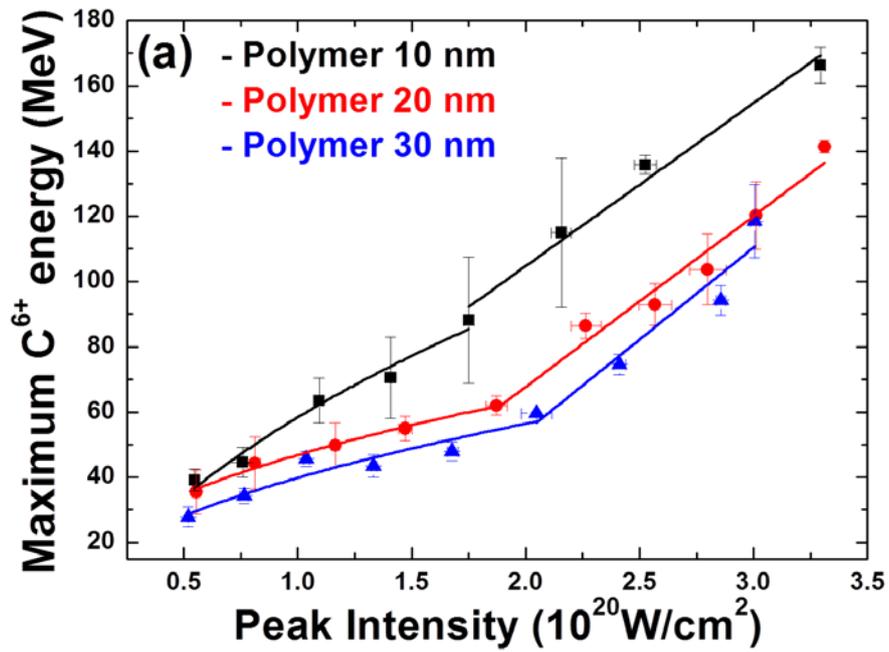

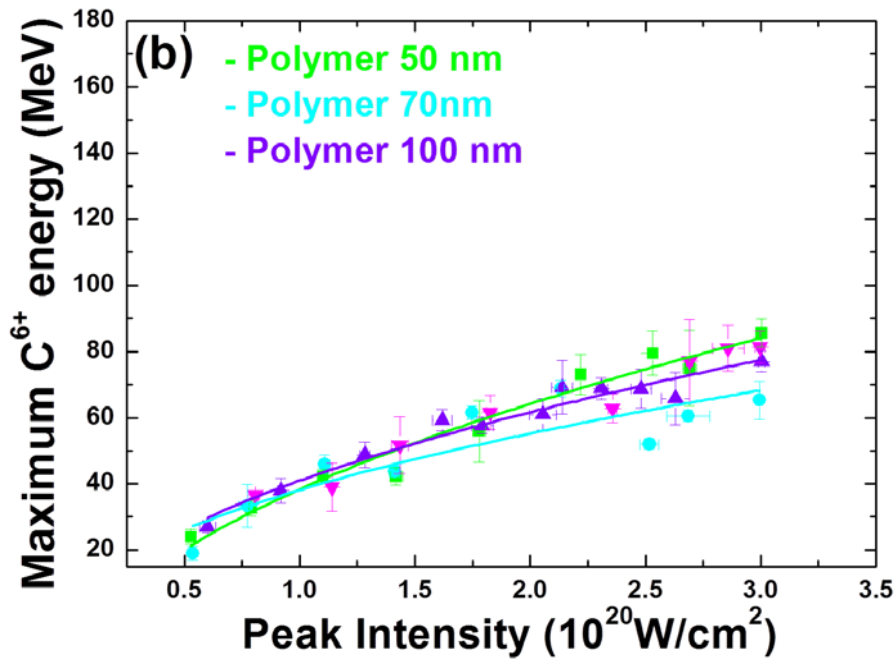

**Supplementary Figure S5: Maximum $C^{6+}$ energy as a function of the laser peak intensity for linearly polarized laser pulse.** (a) Maximum $C^{6+}$ energy for 10-, 20-, and 30-nm-thick polymer target. (b) Maximum $C^{6+}$ energy for 50-, 70- and 100-nm-thick polymer target.

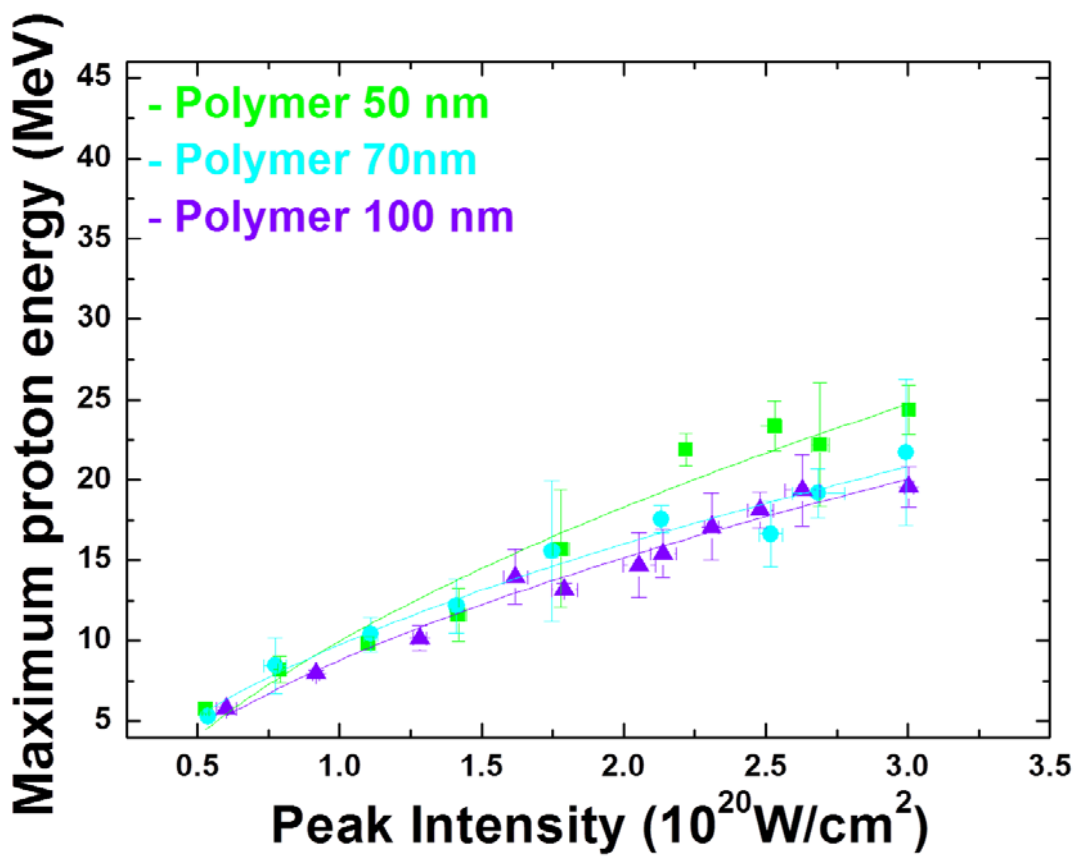

**Supplementary Figure S6: Maximum proton energy as a function of the laser peak intensity for linearly polarized laser pulse.** (a) Maximum proton energy for 50-, 70-, and 100-nm-thick polymer target.

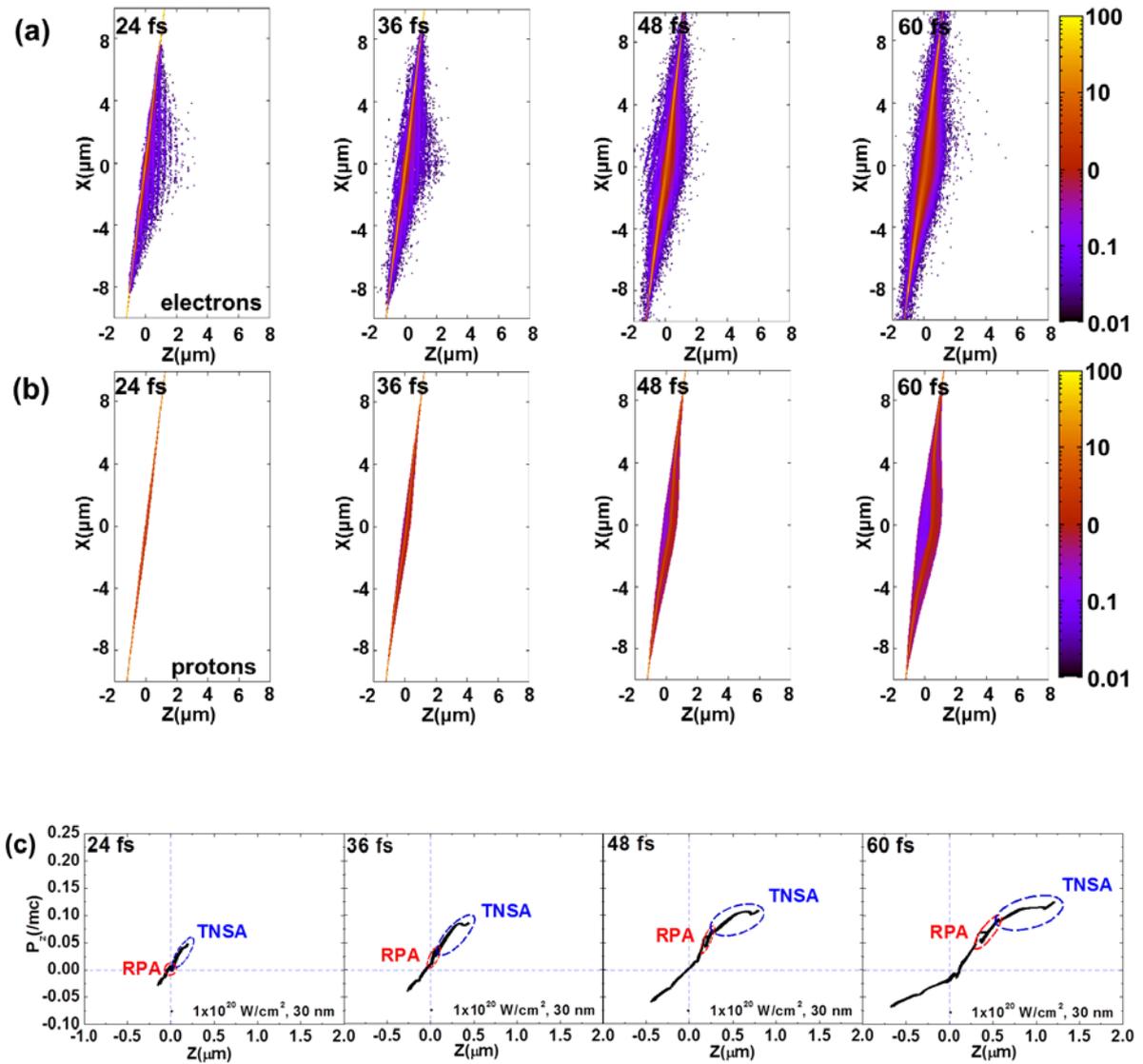

**Supplementary Figure S7: PIC simulation.** Temporal evolution of number density for (a) electrons and (b) protons and (c) proton phase space distribution at t = 24, 36, 48, and 60 fs from a 30-nm-thick target when irradiated by an intensity of $1.0 \times 10^{20}$ W/cm$^2$. The acceleration procedure is similar with that of high intensity case (see Fig. 5). However, as is clearly seen in (c), due to the low intensity, the energy of the RPA-accelerated protons from the target front surface is lower than that of TNSA-accelerated protons from the target rear surface. The dominance of the TNSA mechanism is also confirmed from the Fig. 5(d), where the longitudinal electric field for the lower intensity case (blue line) shows the typical form of the TNSA mechanism. As a result, the maximum proton energy is determined by the TNSA mechanism, which results in the scaling of $E_{max} \sim I^{1/2}$.